\documentclass[12pt,preprint]{aastex}
\usepackage{graphicx,natbib,amsfonts,amssymb,lscape}
\usepackage[figuresright]{rotating}
\usepackage{apjfonts}
\usepackage{color}
\usepackage{lscape}     
\usepackage{graphicx}
\usepackage{graphics}
\usepackage{psfig}
\usepackage{multicol}
\usepackage{natbib}
\usepackage{amsmath}
\newcommand{\A}{\text{\AA}}

\usepackage{graphicx,enumitem}

\usepackage{xcolor}
\usepackage{hyperref}
\hypersetup{
   colorlinks,
   linkcolor={blue!88!black!80},
   citecolor={blue!88!black!80},
   urlcolor={blue!88!black!80}
}

\slugcomment{Accepted to ApJ, on \today}

\usepackage{CJK}
\usepackage[T1]{fontenc}

\begin{document}
\begin{CJK}{UTF8}{gbsn}

\title{HOW FAR IS QUASAR UV/OPTICAL VARIABILITY FROM A DAMPED RANDOM WALK AT LOW FREQUENCY?  }
\author{Hengxiao Guo (郭恒潇) $^{1,2,3,4}$, Junxian Wang (王俊贤)$^{1,2}$, Zhengyi Cai (蔡振翼)$^{1,2}$, Mouyuan Sun (孙谋远)$^{1,2}$\\
$^{1}$CAS Key Laboratory for Research in Galaxies and Cosmology, Department of Astronomy, University of Science and Technology of China, Hefei 230026, China\\
$^{2}$School of Astronomy and Space Science, University of Science and Technology of China, Hefei 230026, China\\
$^{3}$National Center for Supercomputing Applications, University of Illinois, 1205 West Clark St., Urbana, IL 61801, USA\\
$^{4}$Department of Astronomy, University of Illinois, 1002 W. Green Street, Urbana, IL 61801, USA\\
hengxiaoguo@gmail.com, jxw@ustc.edu.cn\\
} \shorttitle{QUASAR VARIABILITY FROM THE DRW MODEL} \shortauthors{Guo, et al.}

\begin{abstract}
Studies have shown that UV/optical light curves of quasars can be described with the prevalent damped random walk (DRW,  also known as Ornstein-Uhlenbeck process) model. A white noise power spectral density (PSD) is expected at low frequency in this model, however, direct observational constraint to the low frequency PSD slope is hard due to limited lengths of the light curves available. Meanwhile, quasars show too large scatter in their DRW parameters to be attributed to the uncertainties in the measurements and the dependence of variation to known physical factors. 
In this work we present simulations showing that, if the low frequency PSD deviates from DRW, the red noise leakage can naturally produce large scatter in variation parameters measured from simulated light curves. The steeper the low frequency PSD slope is, the larger scatter we expect. Based on the observations of SDSS Stripe 82 quasars, we find the low frequency PSD slope should be no steeper than -1.3. The actual slope could be flatter, which consequently requires that quasar variabilities should be influenced by other unknown factors. We speculate that magnetic field and/or metallicity could be such additional factors. 
\end{abstract}

\keywords{galaxies: active -- quasars: general }

\section{INTRODUCTION}
Ubiquitous aperiodic variabilities of quasars can be utilized to trace the physical information in different scales, especially for the inner accretion disk around the supermassive black hole.
With the coming era of time domain survey (e.g., OGLE, \citealt{udalski97}; SDSS Stripe 82, \citealt{sesar07}; CSS, \citealt{drake09}; PTF, \citealt{law09}; TDSS, \citealt{morganson15}; Pan-STARRS, \citealt{kaiser02}; DES, \citealt{des16}; LSST, $~$ \citealt{ivezic08}), quasar variability is gradually attracting more attention.

Measuring the power spectral density (PSD) shapes is one of the major aims of variation studies. An early X-ray timing analysis of a Galactic black hole X-ray binary (BHB) - Cyg X-1 shows that the PSD in the low state is flat below the low frequency 'knee',  follows a slope $\sim$ -1 up to another high frequency 'knee', and then steepens to a slope $\sim$ -2. While in the high state, only one 'knee' has been detected so far, and the slope below and above the break is -1 and around -2 \citep{belloni90,reig02,mchardy06,kelly11}. Most X-ray power spectra of AGNs are similar to those of high state BHB systems, and always only one break is detected \citep{lawrence87,mchardy87,papadakis02,markowitz03}:  
the slope above the break frequency is suggested to be around -2, and flattens to a shallower slope -1 at low frequencies \citep{uttley02,mchardy04,arevalo08b}. The intrinsic mechanism of the break frequencies is likely to be the characteristic timescales (e.g. dynamical timescale or thermal timescale) ending at the edge of inner disk region \citep{mchardy10}.   

In UV/optical bands, the story is different as the characteristic break frequency appears at much lower than that in X-ray.
Recently, \cite{kelly09} proposed that the quasar optical light curves can be well modeled by the DRW model, also known as Ornstein-Uhlenbeck (OU) process \citep{UO30}, where   a self-correcting term is added to a divergent random walk model to push any deviations back toward the mean value. The DRW light curves could be converted to a PSD with slopes $\alpha_{l}$ = 0 and $\alpha_{h}$ = -2 at low and high frequencies, respectively, 
and the break frequency could be associated with the thermal timescale in the accretion disk \citep{kelly09,sun15}. Various investigations confirm that this model could describe not only the UV/optical light curves of quasars \citep{koz10a,mac10,zu11}, but also blazars \citep{ruan12,sobolewska14}. 
Subsequently, DRW process has been widely adopted in literature to interpret quasar variabilities \citep[e.g.][]{charisi16,caplar16,andrae13,kelly14,koz16a,koz16b,koz16c,koz17a,koz17b}. 
Note $\it Kelper$ results with $\sim$ 30 min sampling revealed steeper power spectral slopes of -3 at very high frequency, which significantly deviate from the DRW model \citep{mushotzky11,kasliwal15a}. \cite{zu13} concluded that while on very short timescale (below a few months) there is deviation from the DRW model, on timescale from several months to a few years, the DRW model is well consistent with the observed light curves, as seen with SDSS Stripe 82 data \citep{mac10,mac12} and well sampled OGLE data \citep{zu11,zu13}. 

However at very long timescales (above a few years), it is yet unclear whether there is significant deviation from DRW, i.e., whether the very low frequency PSD has a slope different from 0.
The difficulty mainly comes from the limited length of available light curves, which are not sufficiently long to cover the flat part of the PSD. 
Furthermore, \cite{emma10} pointed out that spurious breaks often emerge in the structure function of almost all light curves even though they may contain no intrinsic characteristic timescale \footnote{This effect has little effect on a statistical work. }. There are a few AGNs which have light curves spanning as long as several decades or even a century, including NGC 4151 \citep{guo14} and Mrk 421 \citep{chen14}, suggesting non-flat PSD at very low frequency, but with poor photometric data obtained in early days.   
Additionally, we note that some quasars show dramatic variations in magnitude on very long timescales, including changing-look quasars \citep[e.g.,][]{osterbrock77,cohen86,lamassa15,ruan16,mac16,mcelroy16}.
These suggest
that it is unlikely that the variation amplitude of quasars simply ceases to rise at very long timescales, as predicted with the simple DRW process. 
Statistical constraints to the very low frequency PSD slope of quasars are still lacking. 

Meanwhile, which physical parameters determine quasars variability is still a riddle to date. Previous studies have suggested that quasar variability amplitude increases with decreasing luminosity, rest frame wavelength and Eddington ratio, and increases with increasing black hole mass, while the correlation with redshift is very weak \citep{wills93,giveon99,vanden04,wold07}. 
A fundamental question is that for quasars with the same physical parameters mentioned above, do they have identical variation properties?

\cite{mac08} showed that the structure functions of quasars for fixed luminosity, rest frame wavelength and timescale show very large scatter.  
\cite{bauer11} and \cite{mac10,mac12} further explored this problem and find that quasars at fixed physical parameters show too large scatter in 
$\hat\sigma$ ($\hat\sigma$=SF$_{\infty}/\sqrt\tau$, where SF$_{\infty}$, $\tau$ are the two key parameters of the DRW model), to be attributed to sparse data sampling and photometric uncertainties. 
They suggested the variability properties of quasars have intrinsic scatter driven by unknown factors, in additional to the physical parameter aforementioned.

However, the red noise leakage could also add a significant scatter in the observed variation parameters of quasars.
Red noise leakage refers to the variability power transferring from low to high frequencies by the lobes of the window function, i.e., short light curves display long term trend of the variation. 
It distorts the measurement of PSD and variation amplitude using light curves with limited length.
The significance of such leakage depends both on the PSD slope and the length of observed light curves \citep{vau03}.
Conversely, the observed scatter in the variation parameters can be used to independently constrain the quasar PSD slope at very low frequency, extending beyond the coverage of available light curves. 

In this paper, we focus on the variation parameter $\hat\sigma$ ($\hat\sigma$=SF$_{\infty}/\sqrt\tau$), which can be much better constrained with observed light curves compared with SF$_{\infty}$ and $\tau$,
to restrain the quasar PSD profile at very low frequency. In \S2, 
we build a sample of quasars from SDSS Stripe 82, for which SDSS $r$ band light curves are modeled with DRW process to derive their $\hat\sigma$.
We show the scatter in the measured $\hat\sigma$ of the sample (after correcting its dependence to known physical parameters, including luminosity, black hole mass, redshift, etc)
is too large to be recovered with simulated DRW light curves (applying the same data sampling and photometric uncertainties from the real data).
In \S3, we simulate light curves using non-DRW models with non-flat low frequency PSD slopes. We show that red noise leakage is able to reproduce the observed scatter,
which in turn yields that the low frequency slope in PSD should be no steeper than -1.3. 
Discussion and conclusions are given in \S 4. 

\section{THE SCATTER OF QUASAR VARIABILITY PARAMETERS}
 
\subsection{SDSS Stripe 82 quasar sample and DRW fitting}

SDSS Stripe 82, lying along the celestial equator in the Southern Galactic Hemisphere, covers an area of $\sim$ 290 square degree, and
has been repeatedly scanned $\sim$ 60 times in the $ugriz$ bands by the SDSS imaging survey. 
\cite{mac12} presented recalibrated $\sim$10 yr long five bands light curves for 9275 spectroscopically confirmed quasars in the field\footnote{$\tt http://www.astro.washington.edu/users/ivezic/macleod/qso\_dr7/$}.
Measurements of black hole mass, absolutely magnitude (K-corrected) and bolometric luminosity for most of them are available, taken from \cite{shen08}.
With the $\sim$ 10 yr long light curves of such a large sample of quasars, we can extensively study the scatter of quasar variability parameters. 

We model the SDSS light curves with DRW process. 
DRW as a stochastic process is well described by the exponential covariance matrix of signal, with its structure function expressed as \citep[e.g.][]{hughes92,mac10}
\begin{equation}
SF(\Delta t)=SF_{\infty}(1-e^{-|\Delta t|/\tau})^{1/2},
\end{equation}
where $\rm SF_{\infty} = \sqrt{2}\sigma$, $\tau$ the characteristic timescale, and $\Delta t$ the time lag. Clearly, 
\begin{equation}
SF(\Delta t \gg \tau)=SF_{\infty}=\sqrt{2}\sigma,
\end{equation}
\begin{equation}
SF(\Delta t \ll \tau)=\sigma \sqrt{\frac{2|\Delta t|}{\tau}}=SF_{\infty}\sqrt{\frac{|\Delta t|}{\tau}}=\hat\sigma \sqrt{\Delta t},
\end{equation}
In reality, measuring $\tau$ and $\rm SF_{\infty}$ is often hard due to limited light curve length.  Instead, $\hat\sigma$ ( = $SF_{\infty}/\sqrt{\tau}$) 
can be measured with much better accuracy, even when $\tau$ and $\rm SF_{\infty}$ are poorly constrained, 
and could be used to as a good proxy to evaluate the properties of quasar variabilities.
According to the continuous time first order autoregressive process \citep[CAR(1),][]{kelly09}, the PSD of a DRW process is given by
\begin{equation}
P(f) = \frac{4\sigma^2\tau}{1+(2\pi \tau f)^2},
\end{equation}
which infers that $P(f) \propto f^{-2}$ when $f > (2\pi \tau)^{-1}$ and $P(f)$ yields to a constant when $f <(2\pi \tau)^{-1}$. In our paper, the nomenclature "DRW" only refers to the CAR(1) process, and "non-DRW" models refer to several special non-CAR(1) models, which are included by a more general model (continuous-time autoregressive moving average, CARMA\citep{kelly14}).

In this work, to make a direct comparison with the results of \cite{mac12}, we select to use the same $r$ band light curves, which have the best photometries among the five bands for the quasars.
We also adopt similar sample selections: (1) only quasars with $\ge$ 40 epochs (to ensure reasonable DRW fitting) and measurements of black hole mass and absolute magnitude are kept;
(2) $\Delta L_{\rm noise} = ln(L_{\rm best}/L_{\rm noise}) > 2 $ and $\Delta L_{\infty} = ln(L_{best}/L_{\infty}) > 0.05 $, where the  $\Delta L_{\rm noise}$, $\Delta L_{\infty}$, $\Delta L_{\rm best}$ are the likelihoods for $\tau$ = 0, $\tau$ = $\infty$ and the best fit $\tau$ \citep {koz10a,zu13}. The former equation is equivalent to setting a noise limit to request the obtained characteristic timescale larger than the average cadence. The latter equation eliminates the fitting results which cannot distinguish the best $\tau$ and $\tau$ = $\infty$ based on the insufficient light curve length.

\begin{figure}
\centering
\includegraphics[width=18cm]{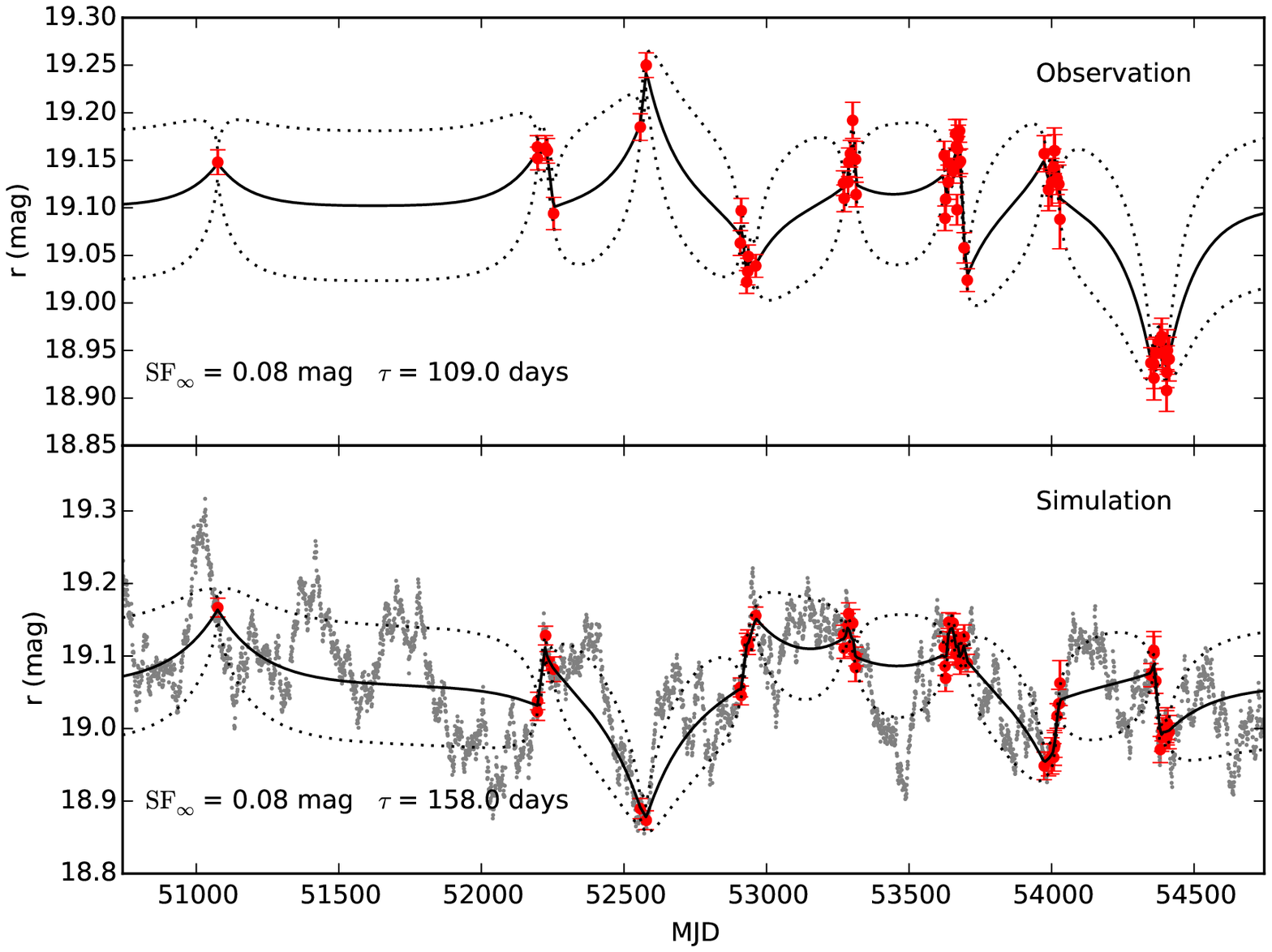}\\
\caption{
Upper panel: $r$ band light curve of a SDSS Stripe 82 quasar, fitted with DRW process. 
Lower panel: Using the DRW parameters obtained in the upper panel (shown in the lower left corner), we make a artificial DRW light curve (grey dots). 
A mock light curve is produced by resampling the simulated DRW light curve and adding photometric errors (red dots and error bars).
The best-fit DRW parameter to the mock light curve is also given.
The solid black lines in both panels are the DRW models with 1$\sigma$ 'error snakes' (black dashed lines).
\label{fig:exam}}
\end{figure}

\begin{figure}
\centering
\includegraphics[width=18cm]{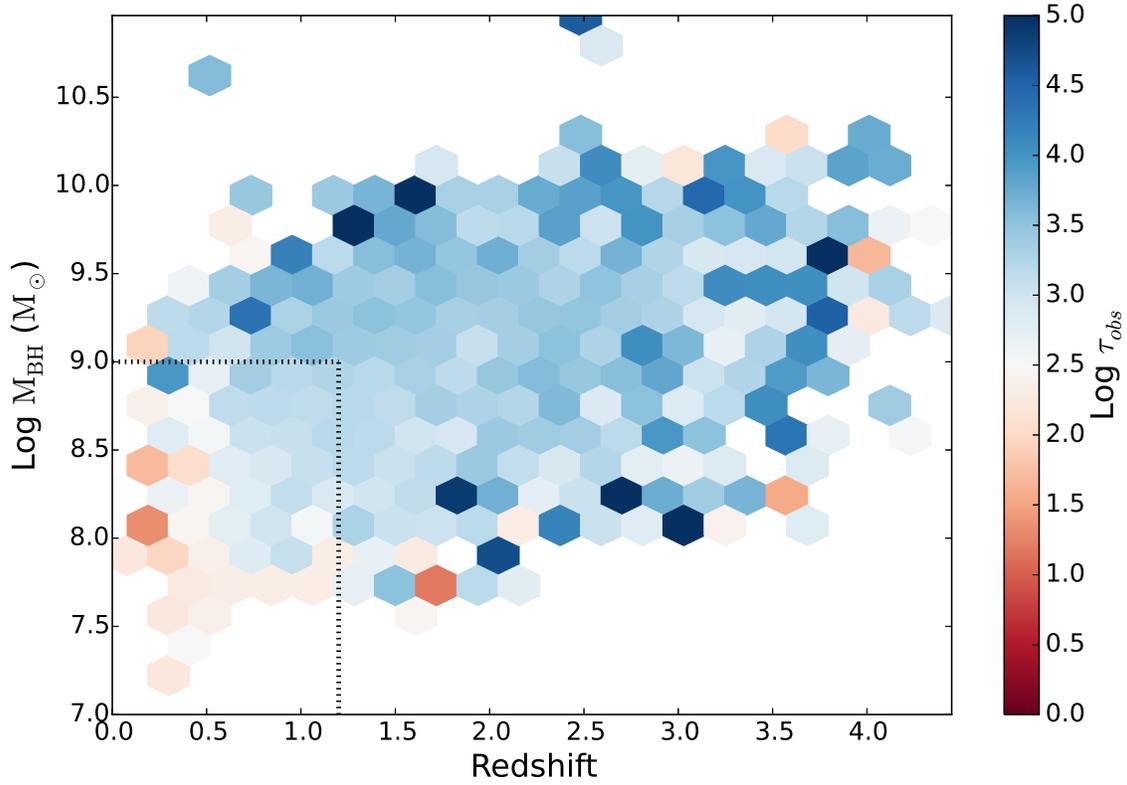}\\
\caption{The color coding distribution of redshift, black hole mass and observed $\tau_{obs}$ for all Stripe 82 quasars. 
Quasars with small $\tau_{obs}$ are mainly located at the lower redshift and black hole mass corner. 
Only quasars within the dotted box are included in this study, for which $\tau$ can be relatively reliably measured 
with SDSS Stripe 82 light curves.
\label{fig:selection}}
\end{figure}

We fit SDSS $r$ band light curves with the Javelin code\footnote{$\tt https://bitbucket.org/nye17/javelin$} \citep{zu11} to measure the damping timescale $\tau$ and asymptotic magnitude $\rm SF_{\infty}$
\footnote{Note we assume a uniform prior on initial input parameters since this non-informative prior is more justicial for unknown parameters, while some other studies favor logarithmic \citep{mac10} or log-uniform prior \citep{charisi16}.} (see the upper panel of Fig. \ref{fig:exam} for a demonstration). 
According to the simulations of \cite{koz17a}, only if the characteristic timescale (in the observed frame, the same hereafter unless otherwise stated) is less than $\sim$ 10\% of the light curve length, the input DRW parameters $\tau_{in}$  can be reasonably recovered by DRW fitting, though biased depending on different priors and fitting method.  
To address this issue, in this study we only include quasars at lower redshift ($ z < 1.2 $) and with low black hole mass (Log $M_{\rm BH} < 9$, see Fig.  \ref{fig:selection}) for which the damping timescale is considerably small comparing with
the light curve length \citep[e.g.][]{koz16c}.
The final sample include 1678 quasars. The median value of their observed $\tau_{obs}$ is 301 days, 
and the typical length of Stripe 82 light curves is $\sim$ 2800 days. The ratio of the typical damping timescale to the light curve duration is 11\%.

\subsection{The scatter of $\hat\sigma$}

\begin{figure}
\centering
\includegraphics[width=18cm]{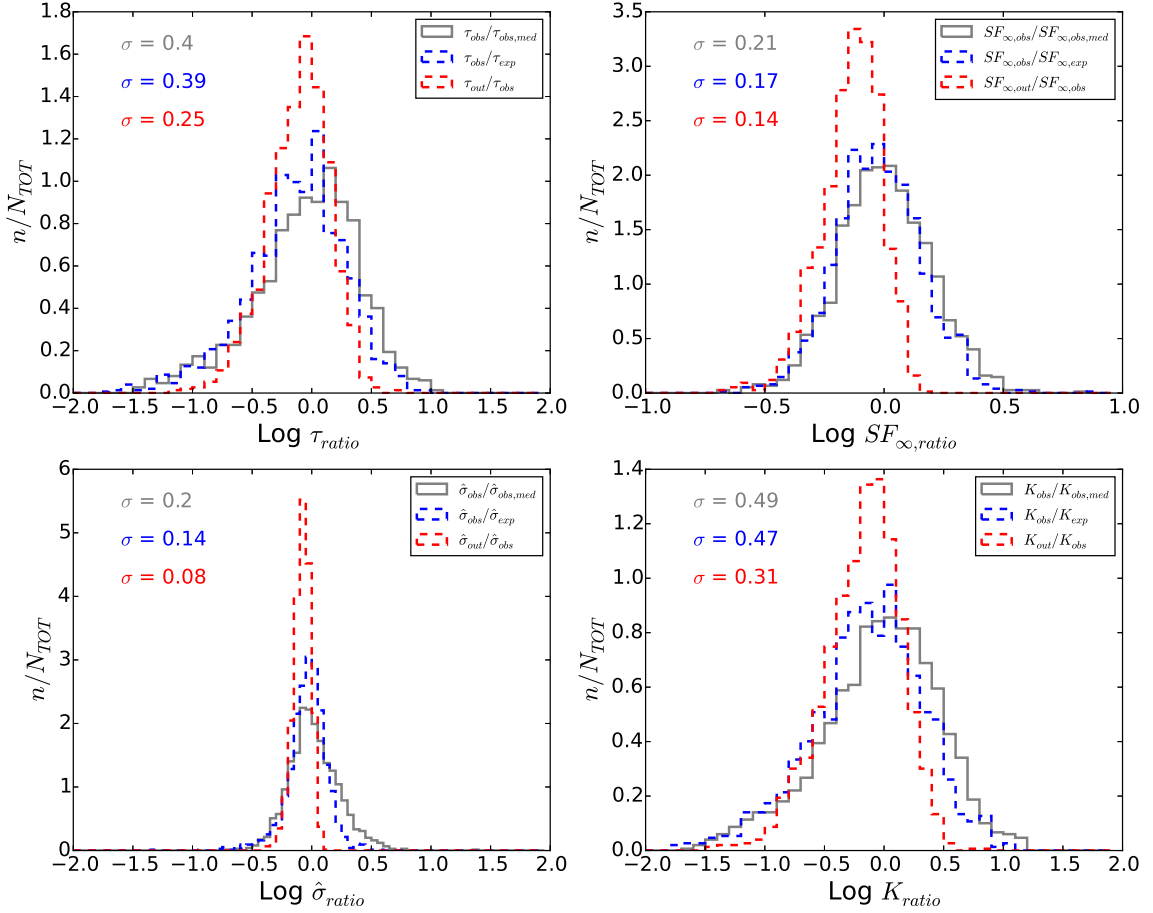}\\
\caption{
The distributions of $\tau_{ratio}$, $SF_{ratio}$,  $\hat\sigma_{ratio}$, $K_{ratio}$. 
Solid grey lines plot the scatter of the observed sample. 
The residual scatters after taking account of dependence to physical factors (Equation \ref{eq:emp}) are plotted as blue dashed lines.
The uncertainties of DRW fitting (due to limited length of light curves, sparse sampling and photometric errors) are plotted as red dashed lines.
The width of distributions are measured with inter-quartile range, and $\sigma$ are given in the upper left corners. 
\label{fig:four}}
\end{figure}

In Fig. \ref{fig:four} we plot the distributions of the measured $\tau_{obs}$, $\rm SF_{\infty, obs}$, $\hat\sigma_{obs}$, and K$_{obs}$ for our quasar sample, respectively, where K = $\tau\sqrt{SF_{\infty}}$ is a variable orthogonal to $\hat\sigma$ in log space. All of them show clear scatters,  but slightly smaller than those of \cite{mac10} as we only include quasars with small black hole mass and redshift, for which $\tau$, $\rm SF_{\infty}$ can be more reliably constrained.

We revisited the empirical relations given by \cite {mac10} between the measured DRW parameters ($\tau$, $SF_{\infty}$ and $\hat\sigma$), and physical parameters including i-band absolute magnitude ($M_{\rm i}$), rest frame wavelength ($\lambda_{\rm RF}$), black hole mass($M_{\rm BH}$) and redshift (z).
Following their method, we first determined the rest wavelength dependence of the variability parameters ($\tau$, $\rm SF_{\infty}$ and $\hat\sigma$) using various SDSS bands,
since each individual quasar has other physical parameters fixed.
Note wen only utilize $u,g,r$ bands here, and exclude $i,z$ bands as in which the weaker intrinsic variabilities and longer intrinsic  damping timescales make the measurements of $\tau$ and $\rm SF_{\infty}$ less reliable than in the bluer bands.
 We then use power-laws to fit other physical parameters simultaneously with fixed wavelength coefficients for r band.
\begin{equation}\label{eq:emp}
\rm log~\it f = A + B \rm log (\frac{\lambda_{RF}}{4000 \A}) + C (\it M_{\rm i}+23) + D \rm log (\frac{\it M_{\rm BH}}{10^9  \it M_{\odot}}) + E \rm log (1+z)
\end{equation}
Using our redefined quasar sample, we obtain A = -0.50, B = -0.48, C = 0.13 and D = 0.18  for f = $\rm SF_{\infty}$ (mag),  A = 2.1, B = 0.66, C = -0.19 and D = 0.14 for f = $\tau$  (days, in the rest frame), and  A =-1.72, B = -0.86, C = 0.22, D = 0.12 for f = $\hat\sigma$ ($\rm mag/day^{1/2}$, in the rest frame).  All Es are fixed to 0, assuming the variability has no evolution with redshift. 
The typical errors of these coefficients are 0.05. The correlations imply that not only the long/short term variability, but also the damping timescale strongly depend on the rest frame wavelength, luminosity and black hole mass.
Note  in Equation \ref{eq:emp}, the dependence to $M_{\rm i}$ and $M_{\rm BH}$ could be coupled as both quantities correlate with redshift.
Caution is thus needed while comparing the coefficients with those reported in other studies \citep[e.g.][]{koz16c,koz17b}.
Obtaining decoupled coefficients requires samples spanning larger ranges in luminosity and $M_{\rm BH}$, and is beyond the scope of this work.
For this study the coefficients we obtained using our sample are sufficient to derive 
the residual scatters in DRW parameters (excluding the dependence to physical parameters listed in Equation \ref{eq:emp}, see next paragraph).

We then obtain the expected $\tau_{exp}$, $\rm SF_{\infty, exp}$,  $\hat\sigma_{exp}$, and K$_{exp}$ for each quasar based on the empirical relations.
In Fig. \ref{fig:four}, we plot the distribution of $\tau_{obs}$/$\tau_{exp}$, $\rm SF_{\infty, obs}$/$\rm SF_{\infty, exp}$, $\hat\sigma_{obs}$/$\hat\sigma_{exp}$, and K$_{obs}$/K$_{exp}$ respectively,
where we can see the residual scatters (blue dotted line, refers to the scatter after excluding the dependence to the physical parameters listed in Equation \ref{eq:emp}), in all four quantities are still prominent. Note the uncertainties in the measurement of black hole mass contribute little to these residual scatters.
For instance, based on Equation 5, a 0.3 dex uncertainty in BH mass can produce a scatter of 0.036 in log($\hat\sigma$), which is only 7\% of the residual scatter of $\hat\sigma$ ($0.036^{2}/0.14^{2}$).

The residual scatters could be at least partially attributed to the uncertainties in the measurement of DRW parameters, due to limited light curve length, sparse time sampling and photometric errors.
Following \citet{mac10,mac12}, we perform Monte-Carlo simulations to quantify such uncertainties. 
Using the observed $\tau_{obs}$ and $\rm SF_{\infty,{obs}}$, we generate a 100 yr long artificial DRW light curve spaced every 1 day for each quasar after eliminating the burn-in part \citep{kelly09}.  A 10 yr segment is randomly cut to cross match with the observed light curve, with the Stripe 82 time sampling and photometric errors imposed at the same time (see the lower panel of Fig. \ref{fig:exam}). 
We generate one artificial light curve for each quasar in our sample, then fit the artificial light curves again to obtain the output DRW parameter.
The ratios of the output to input DRW parameters are over-plotted in Fig. \ref{fig:four}, in which we can see that, the uncertainties in the measurement of DRW parameters can only account for 
41\%, 68\%, 33\% and 44\% of the residual scatter ($\sigma_{red}^2/\sigma_{blue}^2$), for $\tau$, $\rm SF_{\infty}$, $\hat\sigma$, and K respectively. 
Below we focus on the $\hat\sigma$ for which DRW fitting to the simulated light curves recovers the input values with the smallest scatter (red line in the lower left panel of Fig. \ref{fig:four})
and the largest fraction of the  residual scatter (67\%) remains unaccounted.

\begin{figure}
\centering
\includegraphics[width=18cm]{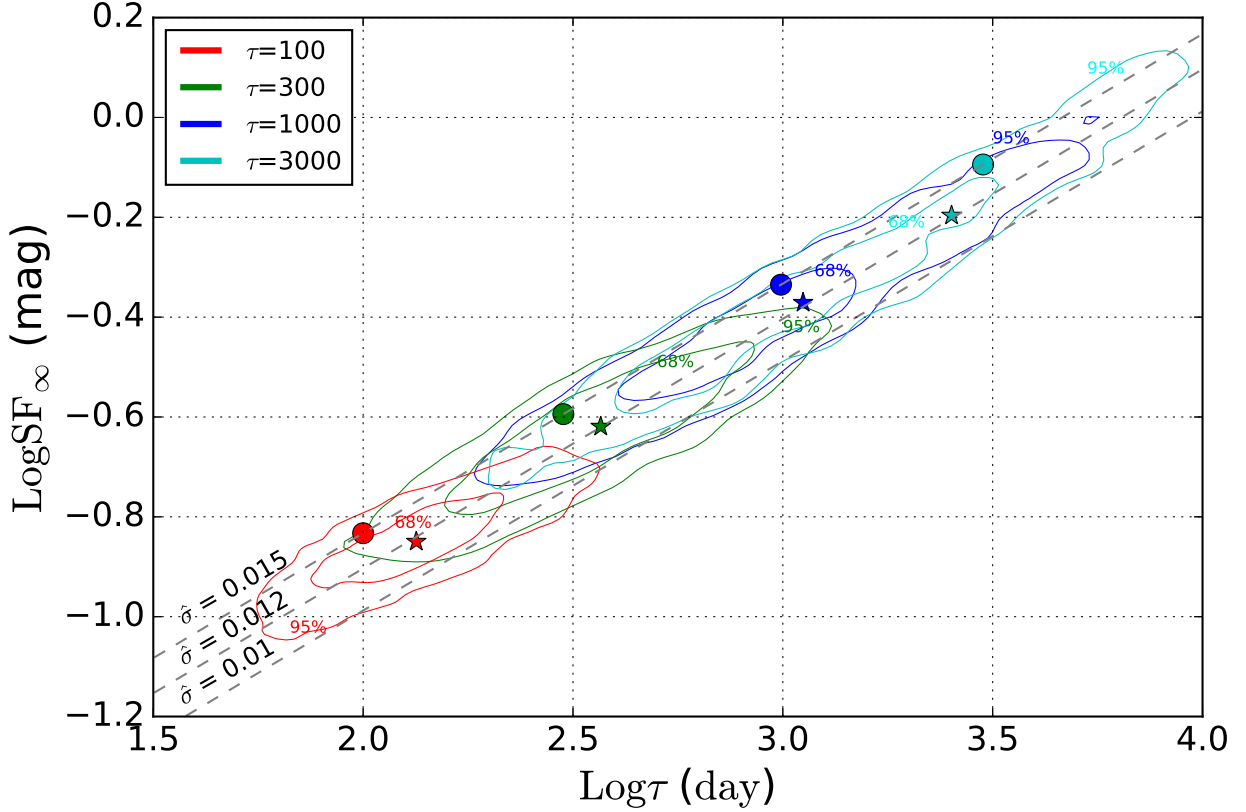}\\
\caption{
The contour distributions of output $\tau - \rm SF_{\infty}$ for four pairs of input values (with constant input $\hat\sigma$). The contours (with red, blue and green cyan corresponding to $\tau_{in}$ = 100, 300, 1000, 3000 days respectively) show that output $\tau$ and $SF_{\infty}$ are tightly coupled, along the direction of constant $\hat\sigma$ (grey dashed lines). The input and median output values are marked with filled circles and stars. With increasing input $\tau$, the scatters of output $\tau$ and $SF_{\infty}$ increase significantly, but that of 
$\hat\sigma$ shows little change. Deviations of the output median values from the input ones (dependent to the sampling and priors adopted in DRW fitting) are seen for $\tau$, $SF_{\infty}$ and $\hat\sigma$. Correcting such small offsets will not alter the conclusions presented in this work. 
\label{fig:samesig}}
\end{figure}

To better demonstrate how the input the DRW parameters were recovered through fitting the artificial light curves, 
in Fig. \ref{fig:samesig} we plot
the contour distributions of output $\tau$ and $\rm SF_{\infty}$ for four different pairs of input values. The input values of $\tau$ are 100, 300, 1000 and 3000 days respectively, and
the input $\rm SF_{\infty}$ was adjusted to keep $\hat\sigma$ unchanged. For each pair of input $\tau$ and $\rm SF_{\infty}$, we generate $\sim$ 2000 artificial light curves and measure output values for each of them.
The contour plots show that the output $\tau$ and $\rm SF_{\infty}$ (for each pair of input values) are clearly coupled, following the direction of constant $\hat\sigma$.
This indicates that the input $\hat\sigma$ can be recovered with remarkably high accuracy and small scatter, comparing with $\tau$ and $\rm SF_{\infty}$ (see also red lines in Fig. \ref{fig:four}).
This holds even for those light curves with intrinsic large $\tau$ ($\sim$ 1000 days), for which the $\tau_{out}$ and $\rm SF_{\infty,out}$ are poorly constrained with huge scatters.
This is also one of the key reasons that we opted to focus on $\hat\sigma$ (but not $\tau_{out}$ or $\rm SF_{\infty}$) in this work.
Offsets of the median output values from input ones are seen for $\tau$, $\rm SF_{\infty}$ and $\hat\sigma$. Such offsets are dependent to the sampling and priors adopted in DRW fitting.
Correcting such offsets will not change the results we present in this work.

The intrinsic $\tau$ for some of the quasars in our sample may still have been significantly underestimated due to limited length of the light curves. 
We perform further simulations to check whether such effect may alter our results.
We enlarge the input $\tau$ value for each source in our sample by a factor of 2 -- 10, and adjust $\rm SF_{\infty}$ accordingly to keep input $\hat\sigma$ unchanged.
We however see no significant difference ($<$ 5\%) in the scatter of $\hat\sigma_{out}/\hat\sigma_{in}$,
demonstrating our results are not affected by possible under-estimations of $\tau$ in our sample.
This can also be seen in Fig. \ref{fig:samesig} that although larger input $\tau$ yield much stronger scatter in $\tau_{out}$,  the scatter of $\hat\sigma_{out}$ appears insensitive to input $\tau$.
We also double check this issue by dividing our sample into two equal-size subsamples according to the black hole mass.
We find that adopting only the lower mass sample (Log $M_{\rm BH}$ $\lesssim$ 8.5), for which the intrinsic $\tau$ should be even smaller,
does not alter the results presented above.

\section{CAN NON-DRW PROCESSES ACCOUNT FOR THE RESIDUAL SCATTER?}

In \S2 we show that the scatter in quasar variability parameter $\hat\sigma$ (modeled with DRW) is too large to be attributed to
uncertainties in DRW parameter measurements (due to limited light curve length, sparse sampling and photometric errors).
Monte-Carlo simulations show such effect can 
only account 33\% of the residual scatter in $\hat\sigma$ (the observed scatter after excluding its dependence on physical parameters in Equation \ref{eq:emp}).
What is the origin of the rest dominant fraction of the scatter in $\hat\sigma$?

\cite{mac10} suggested this could be due to intrinsic scatter in quasar variability controlled by some unknown factors, that is, for quasars with the same black hole mass, luminosity, redshift, and at the same
wavelength,  their variation properties could be distinct. While it is unknown yet what extra factors control quasar variability, in this work we point out that red noise leakage, if quasar variability deviates from DRW at very low frequency, could produce extra scatter in the observed variability parameters. Consequently, we develop an approach to put independent constraint to the low frequency end PSD slope of quasars.

\subsection{Simulating non-DRW Variation}
\begin{figure}
\centering
\includegraphics[]{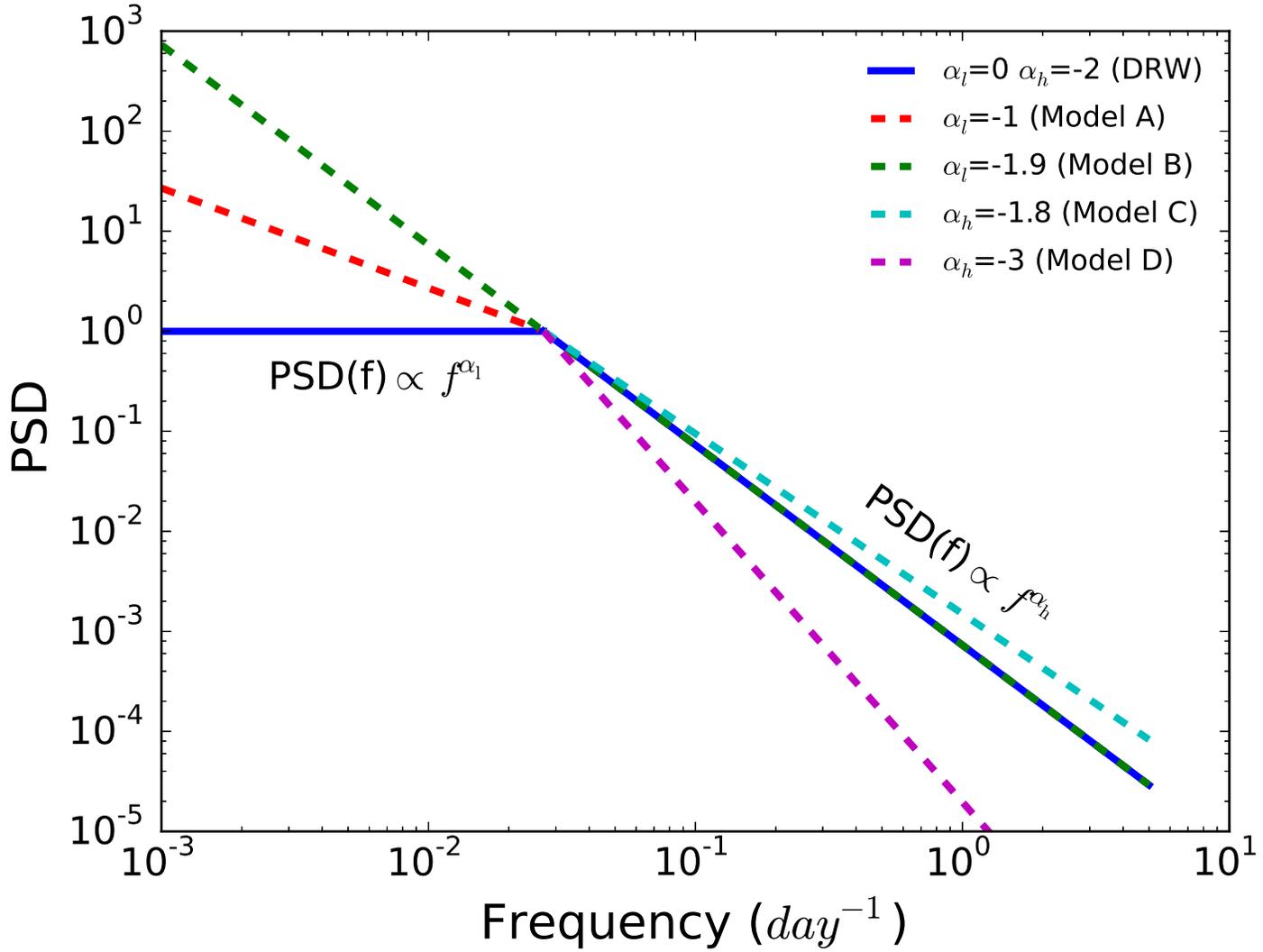}\\
\caption{Demonstration of DRW and non-DRW models. 
\label{fig:demo}}
\end{figure}

Non-DRW light curves are simulated from a broken power-law PSD with public Python code pyLCSIM \footnote{$\tt https://github.com/pabell/pylcsim$}, based on the algorithm of \cite{timmer95}. Two key variables are needed in this approach, one is the break frequency $(2\pi\tau)^{-1}$, where $\tau$ is the damping timescale, another is the rms variability amplitude (rms = $\sigma$ = $\rm SF_{\infty}$ /$\sqrt{2}$ for DRW). 
As a starting point, we use the observed $(2\pi\tau_{obs})^{-1}$ and $\rm SF_{\infty,{obs}}$/$\sqrt{2}$ (where $\tau_{obs}$ and $\rm SF_{\infty,{obs}}$ are the best-fit DRW parameters) of each quasar as input values to generate 100 yr long artificial light curves spaced every 1 day. 
The slopes of the broken power-law model can be arbitrarily decided. 
We investigate four non-DRW models comparing with the standard DRW model:  setting low frequency slopes $\alpha_{l}$ = -1 (model A), -1.9 (model B) with fixed high frequency slopes at $\alpha_{h}$ = -2; and $\alpha_{h}$ = -1.8 (model C), -3 (model D) with fixed $\alpha_{l}$ = 0, respectively (see Figure \ref{fig:demo}). 

\begin{figure}
\centering
\includegraphics[width=18cm]{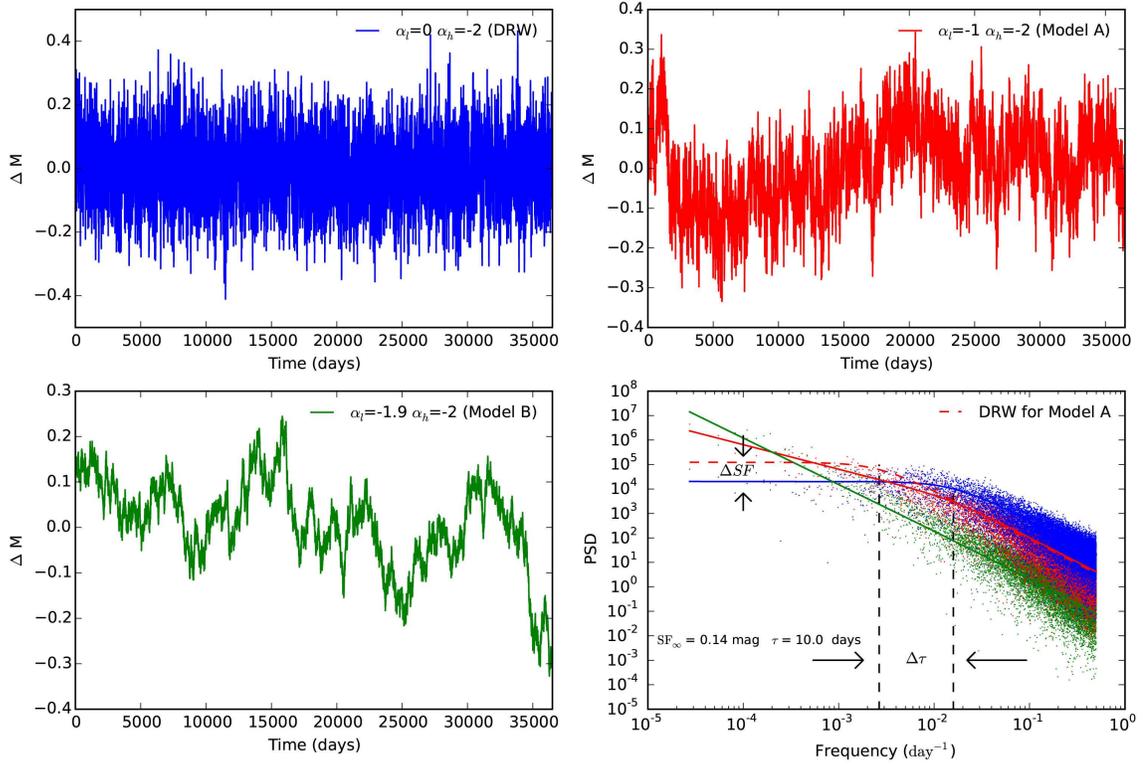}\\
\caption{Examples of our simulated DRW and non-DRW light curves (100 yr long, with $\rm SF_{\infty, in}$ = 0.14 mag and $\tau_{in}$ = 10 days). Their corresponding power density spectra are plotted in the last panel (solid lines), where 
the colored dots are the predicted power spectra with random phase and Poisson noise. 
The red dashed line shows the systematic bias in fitting non-DRW variation (red solid line) with DRW. 
Two vertical black dashed lines mark the break frequencies of the red dashed and solid line, and arrows infer the biases ($\Delta \tau$ and $\Delta \rm SF$).  The PSD unit in last panel is arbitrary. 
\label{fig:nondrw012}}
\end{figure}

Figure \ref{fig:nondrw012} briefly illustrates the deviation of non-DRW models from the DRW one. The first three panels show three 100 yr light curves produced by 3 models with the same input values (rms = 0.1 mag, $\tau$ = 10 days ), with $\alpha_{h}$ fixed at -2 and slopes $\alpha_{l}$ = 0 (DRW), -1 (model A), and -1.9 (model B), respectively. 
In the last panel, the corresponding power density spectra are plotted. 
Note the pyLCSIM code generates light curves with $rms$ (measured through the whole light curve) simply fixed at the input value.
Short segments of such light curve have smaller $rms$ for red noise PSD, but with dispersion 
in $rms$ reflecting red noise leakage.  
We find selecting 10 yr segment out of 100 yr long simulated light curve is sufficient to reproduce the effect of red noise leakage (for the PSD we chose), and simulating even longer
light curves is not needed.
For DRW process, the variation is white noise at timescale longer than $\tau$, thus there is no red noise leakage.
On the other hand, utilizing the DRW model to fit the red noise PSD will misestimate the $\tau$ and $SF_{\infty}$, as shown in the last panel of Fig. \ref{fig:nondrw012} (for the case of $\alpha_{l}$ = -1).

\begin{figure}
\centering
\includegraphics[width=18cm]{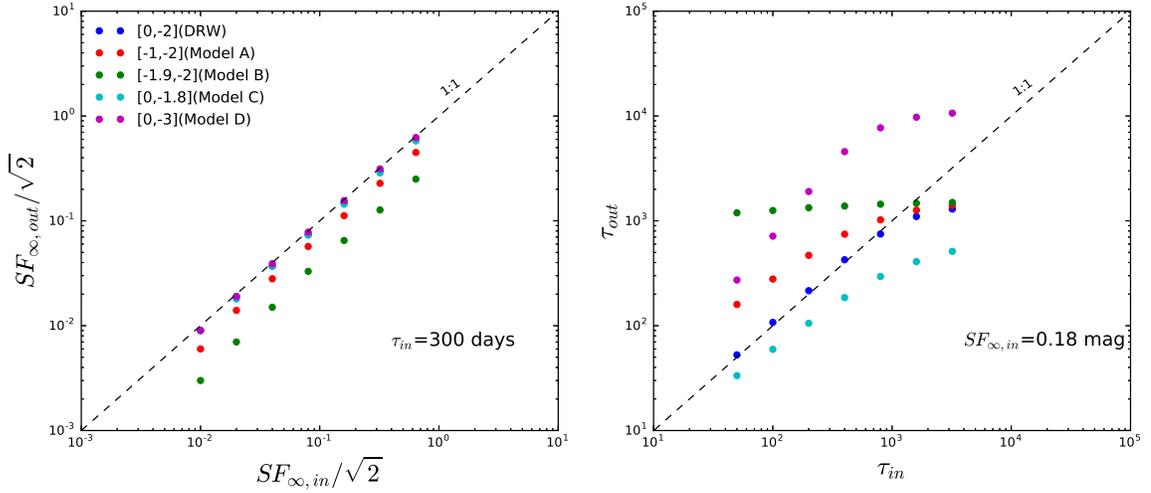}\\
\caption{Best-fit DRW parameters of the simulated light curves versus the input values. 
In the left panel, input $\tau$ is fixed at 300 days with $\rm SF_{\infty}/\sqrt{2}$ ranging from 0.01 to 0.64 mag ($\rm SF_{\infty} = \sqrt{2}\sigma$), while in the the right panel $\rm SF_{\infty}$ is set to 0.18 mag with $\tau$ ranging from 50 to 3200 days. 
\label{fig:offdrw}}
\end{figure}

In Fig. \ref{fig:offdrw} we demonstrate the discrepancies of the output $\tau$ and $\rm SF_{\infty}$ from the input ones used to simulate artificial light curves based on various models. 
For each set of input parameters (shown in Fig. \ref{fig:offdrw} with fixed $\tau$ = 300 days and $\rm SF_{\infty}$ changing from 0.01 to 0.64 mag, or fixed $\rm SF_{\infty}$ = 0.18 mag and $\tau$ changing from 50 to 3200 days), we generate 10000 artificial resampled light curves and plot the averaged output parameters versus the input ones. 
From Fig. \ref{fig:offdrw} we see that for DRW model, the input parameters are well recovered (with only slight offsets), except for that $\tau_{out}$ is significantly underestimated
at $\tau_{in}$ $>$ 1000 days (due to limited length of the light curves), consistent with the results
of \cite{koz17a}. 
For non-DRW models, only model C \& D yield $\rm SF_{\infty, out}$ consistent with input values.
Significant deviations from input $\tau$ are seen for all non-DRW models. 
For model B, particularly, the output $\tau$ barely correlates with the input value. 
This is because the PSD break in model B is so weak that modeling it with DRW process somehow makes no sense (imaging modeling pure red noise PSD slope of -2 with a DRW process).

To correct the biases aforementioned (mismatches between input and output $\tau$ and $\rm SF_{\infty}$), we 
perform simulations for a grid of input parameters ($\tau$ ranging from 1 to 10$^5$ days and $\rm SF_{\infty}$ from 0.01 to 0.8 $\rm mag$). 
The output parameters (averaged over large number of simulations with the same input parameters) build a $\tau_{out}$ -- $\rm SF_{\infty,out}$ surface. 
Finding the position of each quasar on the $\tau_{out}$ -- $\rm SF_{\infty,out}$ plane, we derive the corresponding $\tau_{in}$ and $\rm SF_{\infty,in}$  to be input for simulations.

\subsection{The outputs of non-DRW models}

\begin{figure}
\centering
\includegraphics[width=18cm]{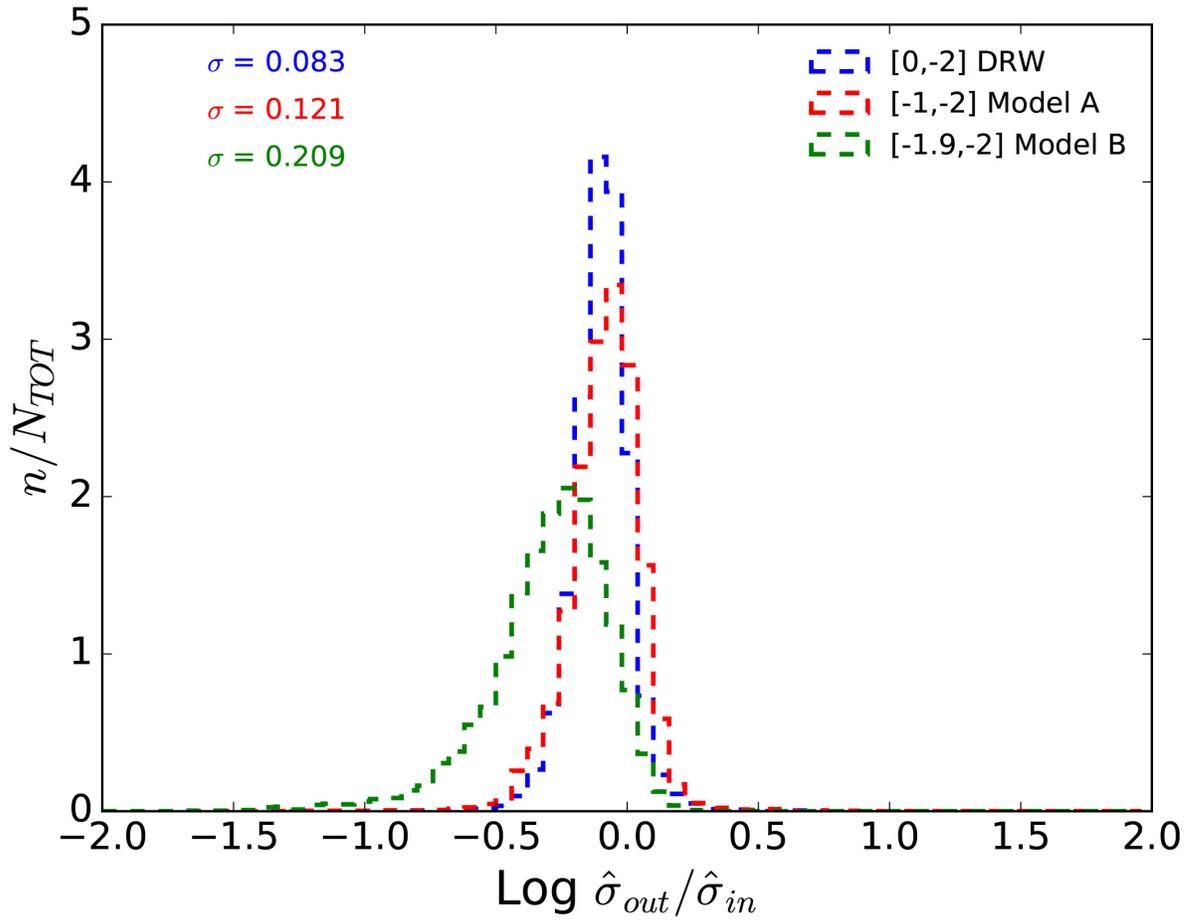}\\
\caption{
The expected scatter in $\hat\sigma$ from our non-DRW models (A \& B), comparing with that from DRW model.
\label{fig:ab}}
\end{figure}

In this section we examine whether non-DRW models could account for the residual scatter in $\hat\sigma$. 
Comparing with DRW, model A \& B (deviation from DRW at low frequency: $\alpha_{l}$ = -1, -1.9 instead of 0, while $\alpha_{h}$ fixed at -2) 
produce obviously larger scatters in $\hat\sigma_{out}$/$\hat\sigma_{in}$ (Fig. \ref{fig:ab}).
This is because of red noise leakage
which refers to the variability power transferred from low to the high frequencies  \citep{uttley02,zhusf16}. 
Such effect is more significant for PSDs with steeper slopes at low frequencies.  
Offset from unity of the $\hat\sigma_{out}$/$\hat\sigma_{in}$ distribution is seen for model B. As explained in \S3.1, this is because 
the PSD break in model B is so weak that modeling it with DRW yields output $\tau$ insensitive to input value, thus our corrections in \S3.1
are not sufficient. Nevertheless, this does not affect the major conclusions of this work.

\begin{figure}
\centering
\includegraphics[width=18cm]{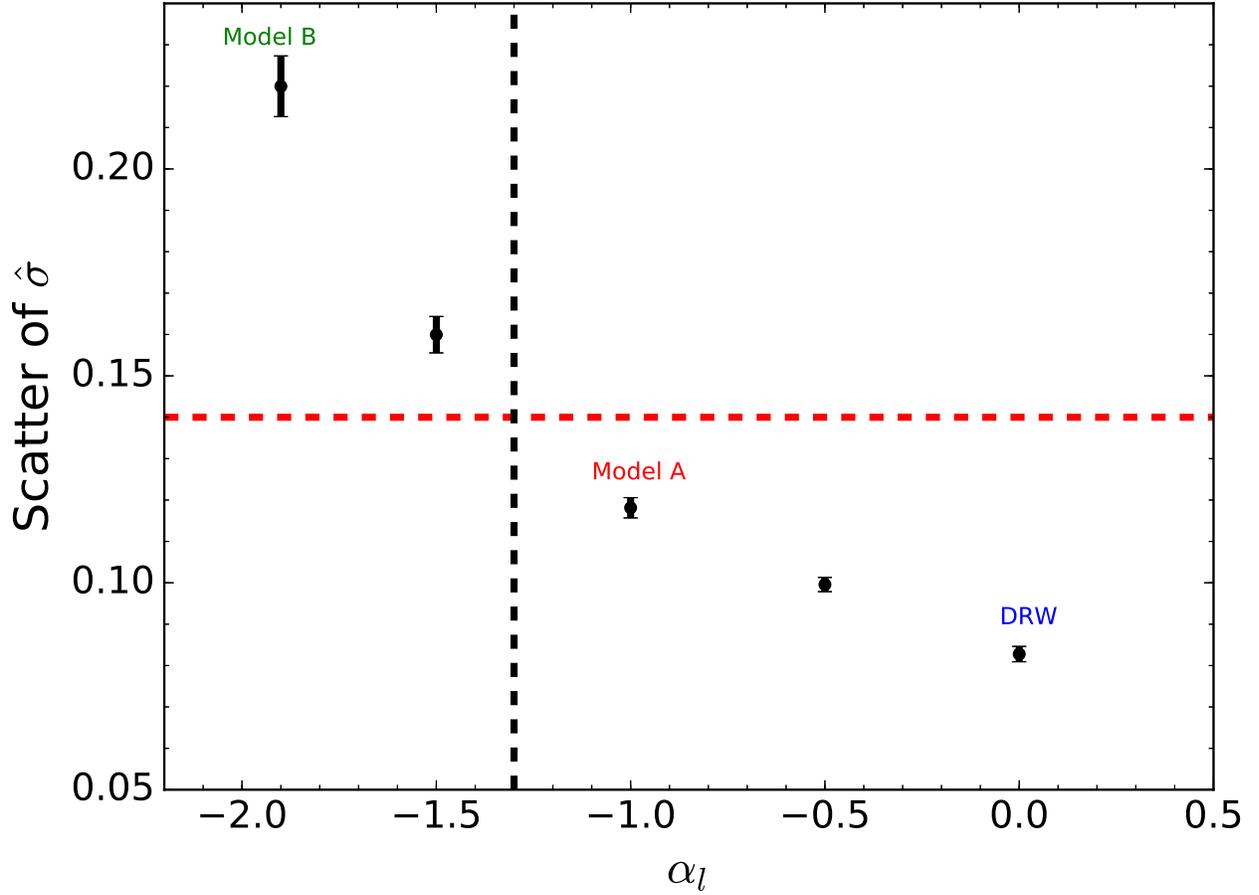}\\
\caption{The expected scatter in $\hat\sigma$ decreases with increasing $\alpha_{l}$ (0, DRW, -1, Model A and -1.9 Model B). The horizontal red dashed line marks the observed residual scatter in SDSS Stripe 82 quasars (see the blue line in the lower left panel of Fig. \ref{fig:four}). The different models are marked with texts.
The DRW model accounts for much lower scatter in $\hat\sigma$ (0.08 $\rm mag/yr^{1/2}$) as compared to empirically measured values (0.14 $\rm mag/yr^{1/2}$, red line). The vertical black dashed line suggests the lower limit of $\alpha_{l}$.
\label{fig:trend}}
\end{figure}

We repeat our simulations for different low frequency slopes.
In Figure \ref{fig:trend}, we plot the reproduced scatter in $\hat\sigma_{out}$/$\hat\sigma_{in}$ versus low frequency PSD slope,  where we see a clear trend that steeper lower frequency PSD yields stronger scatter in $\hat\sigma$. Simulating and modeling quasar light curves with DRW provides a much lower scatter in $\hat\sigma$ (0.08 $\rm mag/yr^{1/2}$)
as compared to empirically measured values (0.14 $\rm mag/yr^{1/2}$).
Non-DRW model with low frequency slope $\alpha_{l} \simeq$ -1.3 appears just sufficient to explain the observed residual scatter in $\hat\sigma$. 

When modeling SDSS stripe 82 light curves, we obtained a run-away fraction (RF) of $\rm RF_{obs}$ = 19\%, i.e., for 19\% of the light curves $\tau$ can not be well constrained (output $\tau$ hits the boundary $\tau$ = $10^{5}$ days). 
The run-away fraction was also measured for our simulations: $11^{+2}_{-2}$ \%  for DRW model, $16^{+2}_{-2}$\% for model A and  $40^{+3}_{-3}$\% for model B,  $6^{+1}_{-1}$\% for model C and  $28^{+2}_{-2}$\% for model D. 
The errors in RF are measured as scatters from different sets of simulations. 
The observed run-away fraction appears more consistent with model A, also suggesting the low frequency PSD slope is likely $\alpha_{l}$ $\sim$ -1, deviated from DRW. 

\begin{figure}
\centering
\includegraphics[width=18cm]{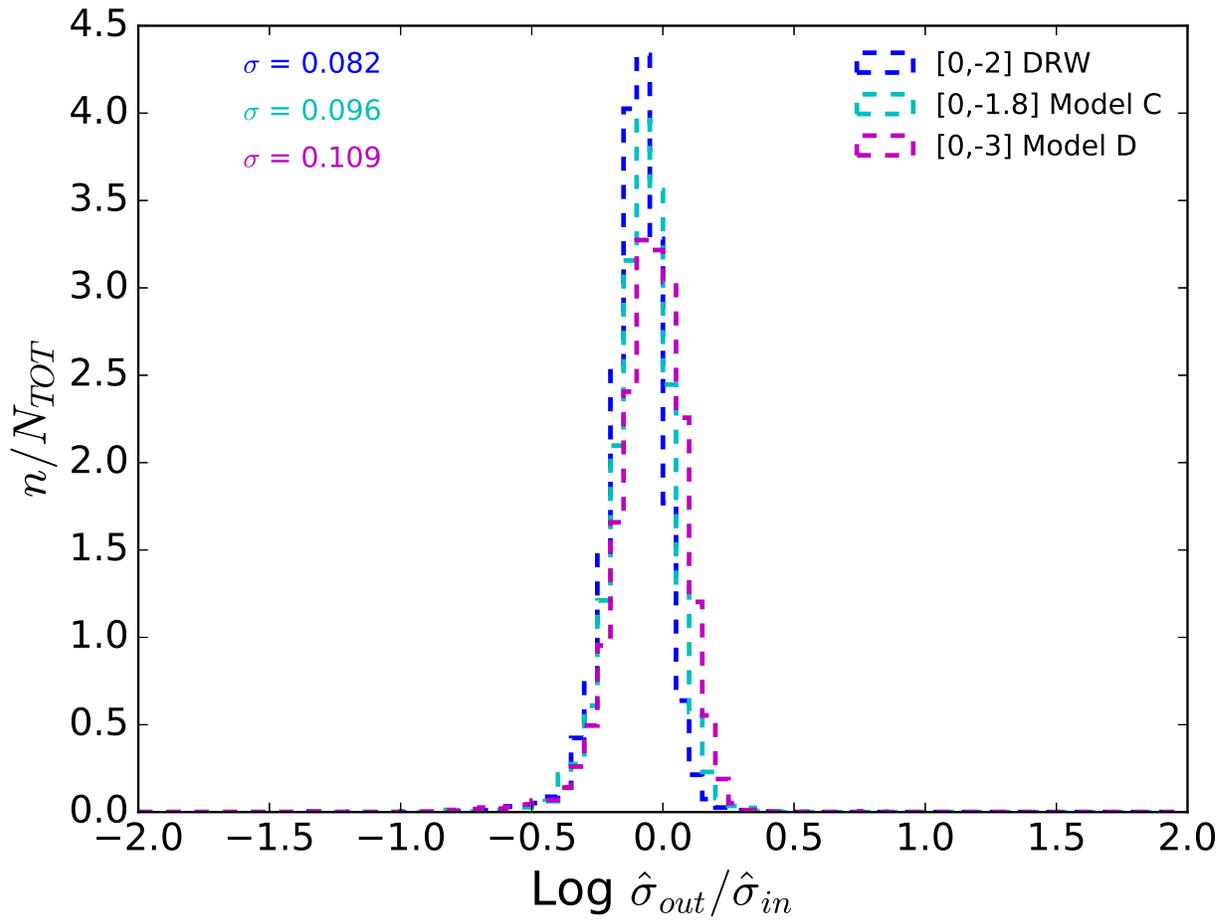}\\
\caption{The same as Fig. \ref{fig:ab}, but for model C \& D. 
\label{fig:cd}}
\end{figure}

For model C \& D ($\alpha_{h}$ = -1.8, -3 instead of -2, while $\alpha_{l}$ fixed at 0),
as there is no red noise leakage, both models yield $\hat\sigma_{out}$/$\hat\sigma_{in}$ scatter similar to that of DRW model (see Figure \ref{fig:cd}).

\section{DISCUSSION AND CONCLUSIONS}

While direct measurement of the low frequency PSD slope is hard for quasars (which requires much longer light curves), our simulations in this work yield a lower limit (below the break, $\alpha_l$ $>$ -1.3). The analyses of the run-away fraction (the fraction of quasar light curves for which the characteristic timescale $\tau$ can not be well constrained) also suggest a red noise low frequency PSD slope ($\alpha_l$ $\sim$ -1.0, also see Macleod et al. 2010)\footnote {However, as we have shown in \S3, the run-away fraction is also sensitive to high frequency PSD slope, thus the constraint to $\alpha_l$ based on run-away fraction analyses alone is rather loose. Contrarily, high frequency slope has little effect to the scatter of $\hat\sigma$.}, instead of white noise predicted by the DRW model. Similarly, low frequency PSD slopes of -1.0 -- -1.2 (on average, with considerably large scatter) were obtained with Pan-STARRS1 light curves \citep{simm16}. 
Deviation from DRW at low frequency can also be seen in the structure function plots presented in Macleod et al. (2012), which includes Palomar archive data to extend the line base up to 50 years for SDSS quasars. 
A red noise low frequency PSD also appears consistent with that seen in X-ray (see Zheng et al. 2017 in preparation for X-ray PSD obtained based on CDF-S light curves spanning over 16 years). 
We note a second break is needed at even lower frequency as the total variability power need to stay finite. 

Long-term variation in quasars could be due to changes of global accretion rate \citep[e.g.][]{li08}, or thermal fluctuation of the disk \citep[see][ for a promising inhomogeneous disk model]{dexter11,cai16}.
 
Low frequency PSD, deviated from single DRW process, could be due to the combination of different DRW processes with various damping timescales \citep{kelly11,dobrotka17}.
Additionally based on the fluctuations of the accretion rate, \cite{lyub97} proposed an accretion disk model with an $\alpha$ parameter fluctuating at different radii, which indicates a roughly flicker-noise type PSD ($P \propto f^{-1 \sim -2}$) with a characteristic timescale approximating the order of the local viscous timescales. Moreover, to account for the intriguing PSD features in Cyg X-1, \cite{take95} proposed a cellular-automaton model for accretion disk based on the concept of the self-organized criticality. In this model , mass accretion takes place in form of avalanches only when the local mass density exceeds critical value, which can produce a $f^{-1.5}$-like PSD. 

Note that a low frequency slope of -1.3 we obtained in this work is a lower limit. If the PSD low frequency slope is actually flatter, an interesting consequence is that there should be other factors (in additional to those aforementioned) which control the variation amplitudes of quasars. We speculate that such factors may include 
metallicity and magnetic field.
Recent 3D radiation magnetohydrodynamic simulations suggest that iron opacity bump may have a strong impact on the thermal stability of accretion disk \citep{jiang16}, pointing to a possible link between metallicity and disk stability. Furthermore, radiative cooling rate increases with increasing metallicity in hot plasma \citep{boe89}, thus inhomogeneous thermal instability may reduce with increasing metallicity in quasars. 
However, direct observational evidence of such link is still absent. 

As to magnetic field,  recent theoretic works have suggest that magnetically driven winds can greatly reduce the disk temperature and help the disk become more stable at a given accretion rate \citep[e.g.][]{li14}. 
Moreover, there are studies claimed that optically thick, geometrically thin accretion disks with strong toroidal magnetic field are stable against thermal and viscous instabilities \citep{begelman07,oda09}. \cite{sadowski16a,sadowski16b} also indicated that magnetic-pressure-dominated thin accretion disks can maintain thermal equilibrium, in contrast to thermally unstable disks dominated by radiation pressure.  
Therefore, a link between magnetic field strength and quasar variability is also expected.

Main conclusions of this work are:

(1) We model 1678 low redshift and low black hole mass quasar light curves of SDSS Stripe 82 quasars with DRW process. Quasars show significant residual scatter in DRW parameters, even after correcting their dependence to known physical parameters, including luminosity, black hole mass, redshift, and rest frame wavelength. 

(2) With simulated DRW light curves, we show that accurately measuring $\tau$ and SF$_{\infty}$ (two key DRW parameters) for an individual quasar using SDSS Stripe 82 light curves is difficult, with huge scatter and strong bias, for sources with intrinsically large $\tau$ ($\gtrsim$ 1000 days in the observed frame), consistent with previous studies.  
However,  $\hat\sigma$ (=SF$_{\infty}/\sqrt\tau$) can be recovered with remarkably small scatter (for $\tau$ up to 3000 days, comparable to the length of the light curves).

(3) The observed residual scatter in $\hat\sigma$ of our quasar sample is too large to be attributed to uncertainties in the DRW parameter measurements caused by limited light curve length, sparse sampling and photometric errors.

(4) We show with simulations that red-noise leakage, if the PSD deviates from DRW below the break frequency,  can produce large residual scatter in $\hat\sigma$.

A low frequency PSD slope of -1.3 is required to match the observed scatter in $\hat\sigma$, indicating the actual low frequency slope should be no steeper than -1.3.

(5) If the actual low frequency PSD slope is flatter than -1.3, there must be other factors (in additional to known ones) which affect the variation amplitudes of quasars, to explain the observed scatter in $\hat\sigma$. Two candidates are magnetic field and metallicity.

\section*{Acknowledgement}
We especially thank the anonymous referee and AAS Journal statistician for their helpful comments and suggestions that have significantly improved the paper. 
We thank Chelsea Macleod, Ying Zu, Riccardo Campana and Xinwu Cao for valuable discussions. This work is supported by the National Basic Research Program of China (973 program, grant No. 2015CB857005), CAS Frontier Science Key Research Program QYCDJ-SSW-SLH006, and National Science Foundation of China (grants No. 11233002, 11421303, and 11503024). J.X.W. is grateful for the support from the Chinese Top-notch Young Talents Program. Z.Y.C. acknowledges support from the Fundamental Research Funds for the Central Universities. M.Y.S. acknowledges support from the China Postdoctoral Science Foundation (grant No. 2016M600485) and the National Science Foundation of China (grant No. 11603022). 

Funding for the Sloan Digital Sky Survey I \&\ II has been provided by
the Alfred P. Sloan Foundation, the U.S. Department of Energy Office of
Science, and the Participating Institutions. SDSS-IV acknowledges
support and resources from the Center for High-Performance Computing at
the University of Utah. The SDSS web site is www.sdss.org.

SDSS is managed by the Astrophysical Research Consortium for the 
Participating Institutions of the SDSS Collaboration including the 
Brazilian Participation Group, the Carnegie Institution for Science, 
Carnegie Mellon University, the Chilean Participation Group, the French Participation Group, Harvard-Smithsonian Center for Astrophysics, 
Instituto de Astrof\'isica de Canarias, The Johns Hopkins University, 
Kavli Institute for the Physics and Mathematics of the Universe (IPMU) / 
University of Tokyo, Lawrence Berkeley National Laboratory, 
Leibniz Institut f\"ur Astrophysik Potsdam (AIP),  
Max-Planck-Institut f\"ur Astronomie (MPIA Heidelberg), 
Max-Planck-Institut f\"ur Astrophysik (MPA Garching), 
Max-Planck-Institut f\"ur Extraterrestrische Physik (MPE), 
National Astronomical Observatories of China, New Mexico State University, 
New York University, University of Notre Dame, 
Observat\'ario Nacional / MCTI, The Ohio State University, 
Pennsylvania State University, Shanghai Astronomical Observatory, 
United Kingdom Participation Group,
Universidad Nacional Aut\'onoma de M\'exico, University of Arizona, 
University of Colorado Boulder, University of Oxford, University of Portsmouth, 
University of Utah, University of Virginia, University of Washington, University of Wisconsin, 
Vanderbilt University, and Yale University.

{}

\end{CJK}
\end{document}